# A Hybrid Artificial-Noise and Secret-Key Scheme for Securing OFDM Transmissions in V2G Networks

Ahmed El Shafie†, Mohamed F. Marzban†, Rakan Chabaan*, Naofal Al-Dhahir†

†University of Texas at Dallas, USA.

*Hyundai-Kia America Tech. Center, Inc.

*Abstract*—We propose a new scheme to enhance the physical-layer security of wireless single-input single-output orthogonal-frequency division-multiplexing (OFDM) transmissions from an electric vehicle, Alice, to the aggregator, Bob, in the presence of an eavesdropper, Eve. To prevent information leakage to Eve, Alice exploits the wireless channel randomness to extract secret key symbols that are used to encrypt some data symbols which are then multiplexed in the frequency domain with the remaining unencrypted data symbols. To secure the unencrypted data symbols, Alice transmits an artificial-noise (AN) signal superimposed over her data signal. We propose a three-level optimization procedure to increase the average secrecy rate of this wiretap channel by optimizing the transmit power allocation between the encrypted data symbols, unencrypted data symbols and the AN symbols. Our numerical results show that the proposed scheme achieves considerable secrecy rate gains compared to the benchmark cases.

## I. INTRODUCTION

The smart grid (SG) enables two-way communications between the utility and the customers (e.g., electric vehicles (EVs), houses). This provides means for data exchange in addition to energy exchange between the utility and consumers. The EVs are connected to the SG for energy charging and discharging, where EVs can buy and sell energy from/to the grid. Although vehicle-to-grid (V2G) connection reduces the power outage possibilities and can help the utility when EVs sell their energies to the grid, the connection introduces new challenges in terms of security since the privacy and security of EVs' owners are vulnerable to eavesdropping. Since an eavesdropping node can extract the identity, activity status, and data of the transmitting devices, it is very crucial to protect the communications systems against such attacks [1], [2].

Securing data transmissions at the physical layer (PHY) using information-theoretic approaches, commonly known as PHY security, was first characterized in the seminal work of Wyner [3]. The secrecy capacity of the link connecting two nodes in a shared medium represents the highest data rate that can be transferred reliably between the legitimate nodes without any information leakage to the eavesdroppers. Hence, in addition to cyber security issues that have been investigated extensively [1], [2], it is critical to investigate the PHY security of wireless V2G communications [4]. In fact, a wireless node with an off-the-shelf radio module can send and receive radio-frequency (RF) signals as well as freely overhear the ongoing communications within radio range. Hence, an adversary, which can be any nearby node, can eavesdrop on the communications between the legitimate EVs and the aggregators or the communications between the aggregators and the utility. This adversary can obtain confidential information about the EVs and the utility. Therefore for a secure EV network design, all potential risks should be considered and new schemes should be designed for secure deployments of wireless V2G networks. Accordingly, in this paper, we propose new robust schemes to ensure secure communications against PHY eavesdropping attacks for wireless single-input single-output (SISO) orthogonal-frequency division-multiplexing (OFDM) transmissions.

Due to its high performance over frequency-selective channels, OFDM transceivers have been widely deployed in both wired and wireless networks. Hence, investigating the security of OFDM systems is crucial. In [5], Li *et al.* investigated the secrecy capacity of OFDM-based wireless networks. To simplify the system's design, Rodrigues *et al.* [6] proposed a low-complexity but insecure metric (from information-theoretic perspective) to study the security of a wiretap channel where the authors defined the system's security as the lower-bound on the minimum mean squared error of the decoded data symbols at the eavesdropping node.

To increase the achievable secrecy rates, spatial artificial-noise (S-AN) injection has been widely investigated in various network settings [7]–[11] while temporal-AN (T-AN) injection for OFDM systems has been recently investigated in [12]. For example, in [8], Tsai and Poor proposed power-allocation schemes for data precoded transmissions aided with precoded S-AN. In [12], Qin *et al.* proposed a new scheme for securing OFDM systems using T-AN for the single-input single-output single-antenna eavesdropper (SISOSE) system. In the T-AN scheme, the cyclic prefix (CP) sequence provides extra temporal degrees of freedom that can be exploited to transmit a precoded time-domain AN vector superimposed over the information vector. The problem was directly extended in [13]. In [14], El Shafie *et al.* proposed a new hybrid S-AN/T-AN scheme and investigated its impact on the average secrecy rate of of the multiple-input multiple-output multi-antenna eavesdropper (MIMOME)-OFDM systems. In [15], the authors revisited the SISO-OFDM system and showed that the number of useful T-AN streams depends on the maximum of the delay spreads of the Alice-Bob and Alice-Eve links.

Another approach for PHY security is based on exploiting the randomness of the wireless channel for secret key generation (SKG). Wireless channels are typically reciprocal at a given carrier frequency. That is, at any instant in time, the multi-path properties are identical in both directions of the link between the transmitter and receiver. Temporal variations of fading channels are a source of randomness that can be measured at both ends of a communication link from which a pair of identical secret keys can be extracted. Secret key extraction and generation from wireless channel measurements has been realized using different properties of the received signal, e.g., received signal strength (RSS) [16], phase differences [17], time delay (in wideband transmission) [18], [19], and channel

This work is supported by Hyundai Inc.

state information (CSI) [20]. Since RSS is easy to measure in practice, it is often used in many scenarios (see, e.g. [16]). Upon agreeing on a secure key between the legitimate partners, it can be used once as a one-time pad (OTP) to encrypt several data symbols. OTP encryption is perfectly secure and provably unbreakable as long as the number of secured data symbols are equal to the number of secret key symbols [21]. Unfortunately, the main restriction of OTP encryption is that the number of secure key symbols is too low [20], [22]–[24] to encrypt all the transmitted data symbols. This is because the channel reciprocity property can not be guaranteed unless Alice and Bob measure the channel simultaneously. Moreover, perfect independence between the Alice-Bob and Alice-Eve links is not possible in some scenarios. Increasing the number of secret key symbols has been investigated in many works, e.g., [20], [22]–[24] and the references therein.

Our main contributions in this paper are summarized as follows. Without Eve's instantaneous CSI knowledge at Alice, we propose a new joint T-AN/SK scheme with a three-level optimization approach to increase the average secrecy rate. The data symbols's power levels are optimized based on the rate of the Alice-Bob channel. In addition, we formulate a constrained optimization problem to maximize the average secrecy rate under a total transmit power constraint at Alice. Hence, we optimize the fraction of transmit power assigned to encrypted data, unencrypted data and AN symbols.

*Notation:* Lower- and upper-case bold letters denote vectors and matrices, respectively. $\mathbf{I}_N$ and $\mathbf{F}$ denote the $N \times N$ identity matrix and the $N \times N$ fast Fourier transform (FFT) matrix, respectively. $\mathbb{C}^{M \times N}$ denotes the set of all complex matrices of size $M \times N$. $(\cdot)^\top$ and $(\cdot)^*$ denote transpose and Hermitian (i.e., complex-conjugate transpose) operations, respectively. $|\cdot|$ cardinality of a set. $\mathbb{R}^{M \times N}$ denotes the set of real matrices of size $M \times N$. $\mathbb{E}\{\cdot\}$ denotes statistical expectation. $[\mathbf{A}]_{k,1:N}$ is the $k$-th row of the matrix $\mathbf{A} \in \mathbb{C}^{M \times N}$. $\mathbf{0}_{M \times N}$ denotes the all-zero matrix with size $M \times N$. $\mathrm{diag} = \{\cdot\}$ denotes a diagonal matrix with the enclosed elements as its diagonal elements. $\mathcal{CN}(x, y)$ denotes a complex circularly-symmetric Gaussian random variable with mean $x$ and variance $y$. $[\cdot]^+ = \max\{0, \cdot\}$ returns the maximum between the argument and zero.

## II. SYSTEM MODEL AND TEMPORAL AN DESIGN

In this section, we state our main assumptions and then explain the design of the proposed AN precoding matrix.

### A. System Model and Assumptions

We investigate the SISO-OFDM wiretap channel scenario where all the wireless nodes are equipped with a single antenna. A legitimate EV (Alice) transmits her confidential messages to a legitimate receiver (i.e., the data aggregator), which we refer to as Bob, in the presence of a passive eavesdropping node (Eve). Eve can be one of the other EVs or an external wireless adversary. We study the possible eavesdropping attacks when an EV communicates with the aggregator. Since Eve is assumed to be passive during the communications between Alice and Bob, her CSI is unknown at Alice. Alice transmits her data using OFDM with $N$ sub-channels. At the beginning of each OFDM block of size $N$ sub-channels, a CP of size $N_{\mathrm{cp}}$ is inserted in the time domain using the CP insertion matrix, denoted by $\mathbf{T}^{\mathrm{cp}} \in \mathbb{R}^{(N+N_{\mathrm{cp}}) \times N}$. Prior to information decoding, the CP is removed at the receiver using the CP removal matrix, denoted by $\mathbf{R}^{\mathrm{cp}} = [\mathbf{0}_{\mathrm{N} \times \mathrm{N}_{\mathrm{cp}}} \mathbf{I}_{\mathrm{N}}] \in \mathbb{R}^{N \times (N+N_{\mathrm{cp}})}$. The main purpose of the CP sequence is to eliminate the inter-block interference and to convert the Toeplitz channel matrix into a circulant one which simplifies the decoding complexity at the receiver. Let $L_{\mathrm{B}}$ and $L_{\mathrm{E}}$ denote the delay spreads (i.e., channel memories) of both Alice-Bob and Alice-Eve links, respectively. We assume a block-fading channel model where the channel coefficients of the Alice-Bob and Alice-Eve links remain unchanged during the coherence time duraion which is sufficient for the transmissions of several OFDM blocks. The channel coefficients change randomly and independently from one coherence time to another. Each tap of the channel impulse response (CIR) of the Alice-Bob and Alice-Eve links is a complex zero-mean circularly-symmetric Gaussian random variable with variance $\sigma^2_{\mathrm{A-B}}$ and $\sigma^2_{\mathrm{A-E}}$, respectively. The CP is designed to be longer than the channel memories of both the Alice-Bob and Alice-Eve links. Let $\mathbf{n}_{\mathrm{B}}$ and $\mathbf{n}_{\mathrm{E}}$ denote the zero mean circularly-symmetric complex-additive Gaussian white noise (AWGN) at Bob and Eve, respectively. Let $\kappa_{\mathrm{B}}$ and $\kappa_{\mathrm{E}}$ denote the noise variances of the receivers at Bob and Eve, respectively. Hence, for a per sub-channel bandwidth of $\Delta_f$, the noise power per sub-channel at Bob and Eve receivers are $\eta_{\mathrm{B}} = \kappa_{\mathrm{B}} \Delta_f$ and $\eta_{\mathrm{E}} = \kappa_{\mathrm{E}} \Delta_f$, respectively.

Let $\mathbf{H}^{\mathrm{time}} \in \mathbb{C}^{(N+N_{\mathrm{cp}}) \times (N+N_{\mathrm{cp}})}$ and $\mathbf{G}^{\mathrm{time}} \in \mathbb{C}^{(N+N_{\mathrm{cp}}) \times (N+N_{\mathrm{cp}})}$ denote the complex time-domain channel matrices of the Alice-Bob and Alice-Eve links, respectively. $\mathbf{H}^{\mathrm{time}}$ is a lower-triangular Toeplitz matrix with $[h^{\mathrm{time}}(0), h^{\mathrm{time}}(1), \ldots, h^{\mathrm{time}}(L_{\mathrm{B}}), 0, \ldots, 0]^\top$ as its first column where $h^{\mathrm{time}}(i)$ is the $i$-th tap of the Alice-Bob link CIR. $\mathbf{G}^{\mathrm{time}}$ can be represented with a similar lower-triangular Toeplitz channel matrix with $(L_{\mathrm{E}} + 1)$ channel taps. Let $\mathbf{H} = \mathbf{F}\mathbf{R}^{\mathrm{cp}}\mathbf{H}^{\mathrm{time}}\mathbf{T}^{\mathrm{cp}}\mathbf{F}^* \in \mathbb{C}^{N \times N}$ and $\mathbf{G} = \mathbf{F}\mathbf{R}^{\mathrm{cp}}\mathbf{G}^{\mathrm{time}}\mathbf{T}^{\mathrm{cp}}\mathbf{F}^* \in \mathbb{C}^{N \times N}$ denote the corresponding frequency-domain complex diagonal channel coefficients of Alice-Bob and Alice-Eve links, respectively. $\mathbf{P}_{\mathrm{x}} = \mathrm{diag}(p_{x_1}, p_{x_2}, \cdots, p_{x_N})$ is a diagonal power matrix containing the transmit power levels at each sub-channel, and $\mathbf{x}$ denotes the $N \times 1$ independent data symbols with zero mean and unit variance.

Our proposed PHY security scheme is as summarized as follows. Alice and Bob use the proposed SKG techniques in [20], [24]. That is, Alice and Bob exchange training signals and estimate their channels. Then, they form a common secret key consisting of a limited number of bits, with which Alice encrypts none, some, or all of the data symbols using an OTP.[1] In general, the secret key rates are low compared to the achievable data rates. Hence, we focus on the case where only a portion of the data symbols (i.e., OFDM sub-channels) can be encrypted using the extracted secret keys. The encrypted symbols are multiplexed in the frequency domain with the remaining unencrypted data symbols. After that, AN symbols are added to the entire OFDM block in the time domain to secure the unencrypted data symbols. Since Bob has the secret key symbols, he can decrypt the encrypted sub-channels and process both encrypted and unencrypted sub-channels. On the other hand, Eve does not have the secret keys to decrypt the encrypted sub-channels and after decoding the unencrypted sub-channels, her signal will be degraded by the T-AN signal.

---

[1]The SKG process is beyond the scope of this paper. SKG algorithms for fading channels under OFDM transmissions are given in [20], [24], where Alice and Bob 1) exchange a known training sequence and measure the channel output, 2) employ error correction techniques to reconcile discrepancies between the measurements/generated secret keys, and 3) perform privacy amplification [25], [26] through one-way hash functions to ensure that Eve does not know the final secret key.

We assume that Alice communicates her data to Bob using a set of constellation points $\mathcal{S}$ that approximates a Gaussian input scheme [26]. Alice and Bob begin by estimating the CIR of their link and agreeing on a secret key $r \in [r_1, r_2, \cdots, r_{N_e}]$, with $r_i \in \mathcal{S}$ using any of the approaches in, e.g., [20], [24] and the references therein. Eve is assumed sufficiently distant from both Alice and Bob such that her probability of obtaining the secret key is very low. Once a secure key has been agreed upon, it can be used as an OTP to encrypt the data transmissions. The OTP key $\mathbf{r}$ ensures a one-to-one mapping from the unencrypted data symbol, $a_i$ to the encrypted data symbol, $\tilde{a}_i$, given the secret key symbol, $r_i$. The security is unbreakable since each secret key symbol $r_i$ is uniformly distributed over $\mathcal{S}$. Hence, no information is leaked to any eavesdropping node that intercepts the encrypted data symbols.

### B. Design of the Temporal AN Precoding Matrix

Since the encrypted sub-channels are secured using keys, the purpose of the T-AN is to secure the unencrypted sub-channels. The T-AN scheme improves the PHY security of the OFDM transmissions by sacrificing a portion of Alice's transmit power to transmit AN vectors to degrade Eve's received signal. The T-AN is injected in the time domain to exploit the temporal degrees of freedom provided by the CP. The precoding matrix, $\mathbf{Q} \in \mathbb{C}^{(N+N_{\text{cp}}) \times N_{\text{cp}}}$, designs the AN to span the null space of the channel matrix between Alice and Bob as follows

$$\mathbf{R}^{\text{cp}}\mathbf{H}^{\text{time}}\mathbf{Q} = 0 \quad (1)$$

The wide matrix $\mathbf{R}^{\text{cp}}\mathbf{H}^{\text{time}} \in \mathbb{C}^{N \times (N+N_{\text{cp}})}$ has a null space of size $N_{\text{cp}}$ columns. Let $\mathbf{z} \sim \mathcal{CN}(0, \mathbf{\Sigma}_z)$ denote the $N_{\text{cp}} \times 1$ AN symbols transmitted to confuse Eve with $N_{\text{cp}}$ streams and let $\mathbf{\Sigma}_z$ denote the AN covariance matrix. The received signals at Bob and Eve can be, respectively, expressed as

$$\mathbf{y}_B = \mathbf{H}\mathbf{P}^{\frac{1}{2}}\mathbf{x} + \mathbf{n}_B \quad (2)$$

$$\mathbf{y}_E = \mathbf{G}\mathbf{P}^{\frac{1}{2}}\mathbf{x} + \mathbf{F}\mathbf{R}^{\text{cp}}\mathbf{G}^{\text{time}}\mathbf{Q}\mathbf{z} + \mathbf{n}_E \quad (3)$$

### III. ACHIEVABLE RATES AND OPTIMIZATION

We propose a three-level optimization technique for the power allocation to increase the average secrecy rate. At the first optimization level, Alice determines the active OFDM sub-channels using the water-filling algorithm as in, e.g., [27] and the references therein. Then, at the second level, Alice optimizes the instantaneous power allocated to each data symbol using the water filling algorithm for each type of data (i.e., encrypted and unencrypted). The reason for these two levels of optimization is that, since Alice does not have any knowledge about Eve's instantaneous CSI, Alice can only optimize her transmit power levels across the OFDM sub-channels to maximize the rate of her own link (i.e. Alice-Bob link). Hence, Alice performs the water-filling algorithm to maximize her transmission rate. After that, at the third level, Alice optimizes the transmit power fraction assigned to AN, encrypted data, and unencrypted data based on the average secrecy rate, which can be computed since Eve's channel statistics are assumed to be known at Alice. Assuming that the total transmit power is $P$, let $\theta = [\theta_1, \theta_2, \theta_3]$ denote the vector of power fractions, where a fraction $\theta_1$ is assigned to encrypted data, a fraction $\theta_2$ is assigned to unencrypted data, and a fraction $\theta_3 = 1 - \theta_1 - \theta_2$ is assigned to AN. This efficient design will lead to a simple 2-dimensional optimization over $\theta_1$ and $\theta_2$ to maximize the average secrecy rate.

We assume that both Bob and Eve know the indices of the encrypted and the unencrypted sub-channels in every OFDM block. Since $\mathbf{G}\mathbf{P}^{\frac{1}{2}}$ is diagonal, Eve can select the sub-channels containing the unencrypted data symbols and decode them jointly. Hence, Eve's received signal after removing the encrypted data sub-channels is given by

$$\mathbf{C}_{\text{ue}}\mathbf{y}_E = \mathbf{C}_{\text{ue}}\mathbf{G}\mathbf{P}^{\frac{1}{2}}\mathbf{x} + \mathbf{C}_{\text{ue}}\mathbf{F}\mathbf{R}^{\text{cp}}\mathbf{G}^{\text{time}}\mathbf{Q}\mathbf{z} + \mathbf{C}_{\text{ue}}\mathbf{n}_E \quad (4)$$

where $\mathbf{C}_{\text{ue}} \in \mathbb{R}^{N_{\text{ue}} \times N}$ is a matrix that extracts the unencrypted OFDM sub-channels whose number is $N_{\text{ue}}$. The power of each AN symbol is $p_z = \frac{\theta_3 P}{L_u}$. Hence, the achievable rate of the Alice-Eve link is given by

$$R_E = \log_2 \det \left( \mathbf{I}_{N_{\text{ue}}} + \mathbf{C}_{\text{ue}}\mathbf{G}\mathbf{P}\mathbf{G}^*\mathbf{C}_{\text{ue}}^* \left( \frac{\theta_3 P \tilde{\mathbf{Q}}\tilde{\mathbf{Q}}^*}{L_u} + \eta_E \mathbf{I}_{N_{\text{ue}}} \right)^{-1} \right) \quad (5)$$

where $\tilde{\mathbf{Q}} = \mathbf{A}_{\text{ue}}\mathbf{F}\mathbf{R}^{\text{cp}}\mathbf{G}^{\text{time}}\mathbf{Q}$ with $\mathbf{A}_{\text{ue}}$ denoting a matrix that extracts the non-zero power unencrypted data sub-channels out of all data sub-channels[2] and $\mathbf{P}$ is the power diagonal matrix of the data symbols.

**Remark 1.** *The expression in (5) assumes that Eve knows the AN precoding matrix $\mathbf{Q}$, which is a best-case scenario for Eve. Given that the null space of a matrix can be obtained from the singular value decomposition (SVD) of that matrix (i.e., by selecting the right singular vectors corresponding to zero singular values), knowledge of the AN precoding matrix $\mathbf{Q}$ may imply that Eve knows some information about the channel matrix between Alice and Bob. This, in turn, might sacrifice the key's security which is based on the CSI of the Alice-Bob link. However, as was shown in Lemma 1 of [26], the right singular vectors of a random matrix do not reveal any information about the matrix itself. Hence, even when Eve knows the AN precoding matrix, she does not know the matrix itself and this information does not reveal additional correlated information about the CSI of the Alice-Bob link. We can therefore assume that Eve has full knowledge of the null space matrix without compromising the information-theoretic security.*

---

[2]The reason of multiplying the interference matrix at Eve, $\mathbf{F}\mathbf{R}^{\text{cp}}\mathbf{G}^{\text{time}}\mathbf{Q}$, by $\mathbf{A}_{\text{ue}}$ is that after performing the power allocation, some of the unencrypted sub-channels will have zero power. Hence, due to the decoupled nature of the OFDM sub-channels, Eve and Bob will not process those sub-channels. Accordingly, Eve needs to drop those sub-channels prior to performing the joint sub-channels (i.e., maximum likelihood (ML)) decoding.

The achievable rate of the Alice-Bob link is given by

$$R_{\rm B} = \sum_{k\in\mathcal{E}}\log_2\left(1+\frac{p_k}{\eta_{\rm B}}|H_k|^2\right) + \sum_{k\in\mathcal{U}}\log_2\left(1+\frac{p_k}{\eta_{\rm B}}|H_k|^2\right) \quad (6)$$

where $\mathcal{E}$ ($\mathcal{U}$) denotes the set of encrypted (unencrypted) sub-channels. The expression in (6) is composed of two terms. The first term is perfectly secured due to encryption using the secret keys while the second term is unsecured and its security should be measured using the secrecy rate. Hence, the instantaneous secrecy rate of the legitimate system is given by

$$R_{\rm sec} = \sum_{k\in\mathcal{E}}\log_2\left(1+\frac{p_k}{\eta_{\rm B}}|H_k|^2\right)$$
$$+ \left[\sum_{k\in\mathcal{U}}\log_2\left(1+\frac{p_k}{\eta_{\rm B}}|H_k|^2\right)\right.$$
$$\left. -\log_2\det\left(\mathbf{I}_{N_{\rm ue}}+\mathbf{C}_{\rm ue}\mathbf{GPG}^*\mathbf{C}_{\rm ue}^*\left(\frac{\theta_3 P}{N_{\rm cp}}\tilde{\mathbf{Q}}\tilde{\mathbf{Q}}^*+\eta_{\rm E}\mathbf{I}_{N_{\rm ue}}\right)^{-1}\right)\right]^+ \quad (7)$$

We notice that, unlike the conventional PHY security schemes, our proposed scheme has a positive secrecy rate due to the use of the secret keys, given by $\sum_{k\in\mathcal{E}}\log_2\left(1+\frac{p_k}{\eta_{\rm B}}|H_k|^2\right)$.

Next, we explain the proposed optimization levels.

### A. Selection of Active Sub-channels

Assuming that the channel gains of the Alice-Bob link in the frequency domain are ordered such that the sub-channel with the highest gain is denoted by $|H_1|^2$ and that with the lowest gain is $|H_N|^2$, we use Algorithm 1 below to determine the active OFDM sub-channels. The algorithm is realized using the total power assigned to data transmission given by $(\theta_1+\theta_2)P$.

**Algorithm 1** Active Sub-channels

1: Initialize $\overline{N}=N$.
2: Order the instantaneous gains of all sub-channels.
3: Compute $C = \frac{(\theta_1+\theta_2)P}{\overline{N}} + \frac{1}{\overline{N}}\sum_{k=1}^{\overline{N}}\frac{\eta_{\rm B}}{|H_k|^2}$.
4: Compute $E_k = C - \frac{\eta_{\rm B}}{|H_k|^2}$; $k=1,2,\cdots,\overline{N}$.
5: If $E_k \leq 0$, remove that sub-channel (should be one with highest index since it has lower gain and SNR).
6: Set $N_{\rm old} = \overline{N}$.
7: Set $N_{\rm new} = \overline{N}$ − number of removed sub-channels.
8: If $N_{\rm new} \neq N_{\rm old}$, $\overline{N} = N_{\rm new}$ and go back to step (3).
9: If $N_{\rm old} = N_{\rm new}$, $|\mathcal{A}| = N_{\rm new}$ and the active set $\mathcal{A}$ has been constructed.

### B. Optimizing the Sub-channels' Power Levels

After determining the set of active sub-channels, denoted by $\mathcal{A}$, Alice encrypts $N_{\rm e}$ sub-channels of them using the available secret keys symbols and the remaining sub-channels are left unencrypted. Assume that the set of encrypted sub-channels is $\mathcal{E}$ and the set of unencrypted sub-channels is $\mathcal{U}$. To obtain the power levels for the encrypted and the unencrypted sub-channels, Alice solves the following two constrained optimization problems,

$$\max_{p_k,\forall k\in\mathcal{E}}: \sum_{k\in\mathcal{E}}\log_2\left(1+\frac{p_k}{\kappa_B}|H_k|^2\right), \text{ s.t. } 0\leq\sum_{k\in\mathcal{U}}p_k\leq\theta_1 P \quad (8)$$

and

$$\max_{p_j,\forall j\in\mathcal{U}}: \sum_{j\in\mathcal{U}}\log_2\left(1+\frac{p_j}{\kappa_B}|H_j|^2\right), \text{ s.t. } 0\leq\sum_{j\in\mathcal{U}}p_j\leq\theta_2 P \quad (9)$$

The optimal power assignment for each optimization problem is a well-known water-filling approach [28].

### C. Maximizing the Average Secrecy Rate

Assuming the knowledge of the statistics of Eve's channel at Alice, the average secrecy rate can be calculated and we can state the following optimization problem to obtain $\theta_1$ and $\theta_2$ that maximize the average secrecy rate

$$\max_{\theta_1,\theta_2}: \mathbb{E}\{R_{\rm sec}\}, \text{ s.t. } \theta_1\geq 0, \theta_2\geq 0, \theta_1+\theta_2\leq 1 \quad (10)$$

A simple method to obtain the optimal solution of the two optimization parameters that maximizes the average secrecy rate is to divide the range of each optimization variable, which spans from 0 to 1, into $M$ equally-spaced grid points, and then evaluate the objective function $\frac{M^2}{2}$ times since we have 2 optimization variables. The $\frac{1}{2}$ term is due to the constraint $\theta_1+\theta_2\leq 1$. The optimal solution is the one that yields the highest objective function in (10). Our proposed approach is summarized in Algorithm 2. Although these computations are performed **offline and infrequently**, their complexity can be reduced by using alternative computationally-efficient methods such as the interior-point method [29, Chapter 11].

**Algorithm 2** Optimization Procedure

1: Select a large number $K$
2: *loop1*:
3: Generate $0\leq\theta_1\leq 1$ and $0\leq\theta_2\leq 1$
4: Compute $\theta_3 = 1-\theta_1-\theta_2$
5: Set $i=1$
6: *loop2*: **while** $i\neq K$
7: Generate channel realizations using the channel statistics
8: For the given channel realization, Solve (8) and (9)
9: Compute $R_{\rm sec}$ using (7)
10: Set $z(i) = R_{\rm sec}$
11: Set $i=i+1$
12: **goto** *loop2*.
13: **endwhile**;
14: Compute the average secrecy rate $\mathbb{E}\{R_{\rm sec}\}$ using $\frac{\sum_{i=1}^{K}z(i)}{K}$
15: **goto** *loop1*.
16: Select $\theta_1$ and $\theta_2$ that maximize $\mathbb{E}\{R_{\rm sec}\}$

### D. Per Sub-channel Processing at Eve

Under per sub-channel processing, Eve decodes each OFDM sub-channel individually. Hence, the received signal at Eve per sub-channel $k$ is given by

$$y_{{\rm E},k} = \mathbf{G}_k\sqrt{p_k}x_k + [\mathbf{FR}^{\rm cp}\mathbf{G}^{\rm time}\mathbf{Q}]_{k,1:N_{\rm cp}}\mathbf{z} + n_{{\rm E},k} \quad (11)$$

where $y_{{\rm E},k}$ is the $k$-th element of $\mathbf{y}_{\rm E}$, $[\mathbf{FR}^{\rm cp}\mathbf{G}^{\rm time}\mathbf{Q}]_{k,1:N_{\rm cp}}$ is the $k$-th row of $\mathbf{FR}^{\rm cp}\mathbf{G}^{\rm time}\mathbf{Q}$, and $n_{{\rm E},k}$ is the $k$-th element of $\mathbf{n}_{\rm E}$. The T-AN power on the $i$-th data sub-channel is given by, $\eta_i = \frac{\theta_3 P}{N_{\rm cp}}[\tilde{\mathbf{Q}}\tilde{\mathbf{Q}}^*]_{i,1:N}$. The achievable rate of the Alice-Eve rate under the per sub-channel processing is given by

$$R_{\rm E} = \sum_{i\in\mathcal{U}}\log_2\left(1+\frac{|G_i|^2 p_i}{\eta_i+\eta_{\rm E}}\right) \quad (12)$$

which suggests that all OFDM sub-channels will be degraded by the AN signal and the correlation of the AN signal across the data sub-channels is not exploited. This degrades Eve's instantaneous rate significantly unlike the joint OFDM sub-channels processing case. The achievable instantaneous secrecy rate is thus given by

$$R_{\text{sec}} = \sum_{k \in \mathcal{E}} \log_2\left(1 + \frac{|H_k|^2 p_k}{\eta_{\text{B}}}\right) \\ + \left[\sum_{i \in \mathcal{U}} \log_2\left(1 + \frac{|H_i|^2 p_i}{\eta_{\text{B}}}\right) - \sum_{i \in \mathcal{U}} \log_2\left(1 + \frac{|G_i|^2 p_i}{\eta_i + \eta_{\text{E}}}\right)\right]^+ \quad (13)$$

IV. SIMULATION RESULTS

The parameters used to generate the figures are: $N = 64$, $L_{\text{B}} = L_{\text{E}} = L = N_{\text{cp}} = 16$, a uniform CIR power delay profile with $\sigma^2_{\text{A-B}} = \sigma^2_{\text{A-E}} = 1/(L+1)$, $\eta_{\text{B}} = \eta_{\text{E}} = \kappa \Delta_f = 0$ dB and SNR = 30 dB per OFDM sub-channel. The average secrecy rates are plotted in bits/sec/Hz. Therefore, all rate-expressions derived in the previous sections are divided by $(N + N_{\text{cp}})$ to convert their units from bits/OFDM block into bits/sec/Hz. Fig. 1 shows the average secrecy rates versus the number of encrypted sub-channels while Eve performs the conventional per sub-channel decoding or the high-complexity joint sub-channels decoding. In the joint decoding case, Eve exploits the T-AN correlation among the OFDM sub-carriers. Hence, the average secrecy rate is reduced significantly for low and moderate numbers of encrypted sub-channels. This shows that the effectiveness of the T-AN-only scheme (corresponding to the case when $N_{\text{e}} = 0$) is reduced significantly when Eve performs joint OFDM sub-channels decoding. The proposed hybrid T-AN/SK-aided scheme enhances the security for both decoding scenarios as $N_{\text{e}}$ increases. In all the following figures, we consider the best-case scenario for Eve where joint sub-channels decoding is performed.

In Fig. 2, we demonstrate that the average secrecy rate is monotonically increasing with the number of encrypted OFDM sub-channels when the transmit power is allocated equally among the data OFDM sub-channels. Three methods are used to select the encrypted OFDM sub-channels. In the first method, we encrypt the $N_{\text{e}}$ sub-channels with the highest instantaneous Bob channel gains among the active sub-channels. In the second method, the $N_{\text{e}}$ sub-channels with the lowest instantaneous Bob channel gains among the active sub-channels are encrypted. In the third method, Alice simply encrypts random $N_{\text{e}}$ sub-channels among the active ones. As depicted in Fig. 2, the three methods achieve similar average secrecy rates. The first method enhances the first term of the secrecy rate in Eqn. (13) (encrypted rate) at the expense of the second term (unencrypted rate) in contrast to the second method. The third method balances between the two terms. That is why, on average, the three methods achieve similar performance.

In Fig. 3, the performance of the T-AN only, SK-aided only and the hybrid T-AN/SK-aided schemes are evaluated. We show that without AN and secret keys, the average secrecy rate is close to zero. This is because the average rates of the Alice-Bob link and the Alice-Eve link are equal. The performance of the T-AN only scheme is close to 1 bit/sec/Hz as Eve performs joint OFDM sub-channels decoding. The SK-aided without the AN scheme has low performance when $N_{\text{e}}$ is small and the performance is enhanced with increasing $N_{\text{e}}$. The hybrid AN/SK-aided scheme combines the benefits of both schemes where the AN impact results in an upward shift and the performance is enhanced with increasing $N_{\text{e}}$ until the point where all the OFDM sub-channels are encrypted and the achievable rate is the Alice-Bob link rate, which is the same as the case of no eavesdropping. Two scenarios are considered for selecting the power fractions in AN/SK-aided scheme. We observe that Algorithm 2 outperforms the fixed power fraction allocation scheme since it optimizes the power levels according to the number of encrypted sub-channels.

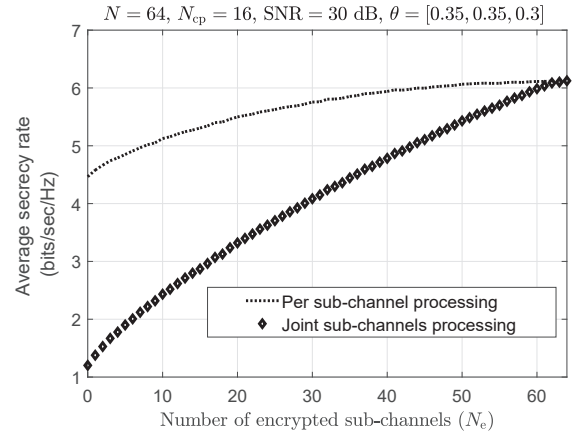

Fig. 1: Achieved average secrecy rate when Eve performs joint and per sub-channel processing.

In Fig. 4, we investigate the impact of $\theta_1$ and its optimal value for a fixed $\theta_3$ on the average secrecy rate. We notice that as $N_{\text{e}}$ increases, the optimal $\theta_1$ moves toward 1. This implies that most of the transmit power will be assigned to the encrypted data. On the other hand, when $N_{\text{e}}$ is small, the optimal values moves toward $1/2$ because it is more secure to assign more power to the unencrypted data which is protected by the AN. However, in general, we have a set of optimal $\theta_1$ (and hence a set of optimal $\theta_2 = 1 - \theta_1 - \theta_3$) which achieves almost the same average secrecy rate.

V. CONCLUSIONS

We showed that using secret keys, generated from the channel randomness, to encrypt some data sub-channels can efficiently increase the average secrecy rate of OFDM transmissions. We showed that the T-AN only scheme performance is degraded significantly when Eve performs joint sub-channels decoding. Our proposed hybrid T-AN/SK-aided scheme enhances the average secrecy rate for both per sub-channel and joint sub-channels decoding strategies at Eve. We proposed a three-level optimization algorithm for optimizing the transmit power distribution between encrypted, unencrypted and T-AN symbols which achieves considerable secrecy rate gains compared to other power allocation strategies. We showed that even without knowledge of Eve's instantaneous CSI, the instantaneous rate of the Alice-Bob channel can be achieved, which is the highest possible secrecy rate, as the number of encrypted data symbols increases.

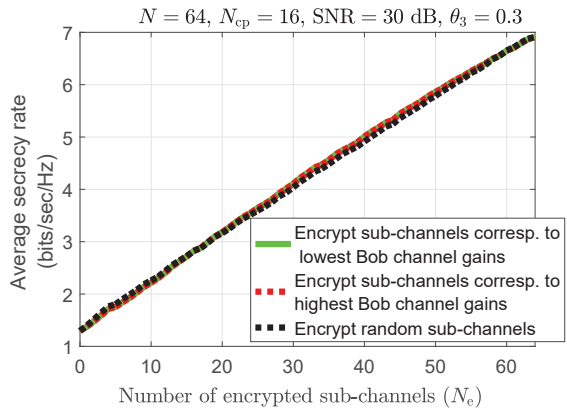

Fig. 2: Average secrecy rate versus number of encrypted OFDM sub-channels.

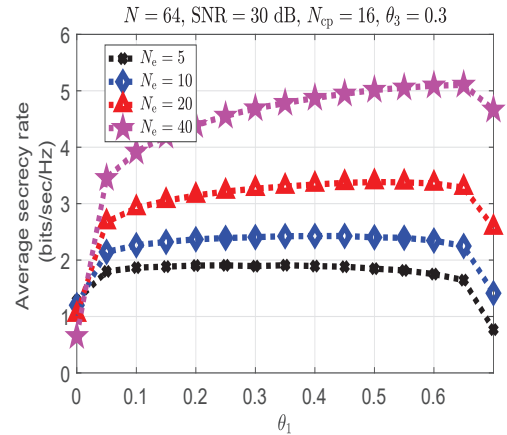

Fig. 4: Average secrecy rate versus number of encrypted OFDM sub-channels.

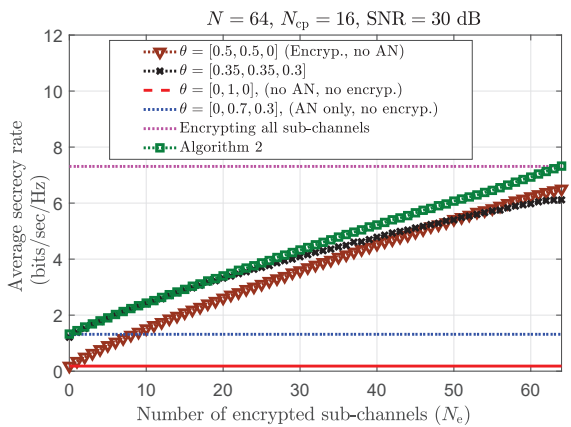

Fig. 3: Average secrecy rate versus number of encrypted OFDM sub-channels.